\documentclass[conference,twocolumn]{IEEEtran}
\usepackage{amsfonts}
\usepackage{times}
\usepackage{graphicx}
\usepackage{latexsym}
\usepackage{dsfont}
\usepackage{amssymb}
\usepackage{amsmath}
\usepackage{cite}
\usepackage{verbatim}
\usepackage{subfigure}

\newcommand{\figref}[1]{{Fig.}~\ref{#1}}

\def\bb0{{\mathbb{0}}}

\def\ba{{\mathbf{a}}}
\def\bb{{\mathbf{b}}}

\def\bff{{\mathbf{f}}}

\def\bh{{\mathbf{h}}}

\def\b0{{\mathbf{0}}}

\def\bX{{\mathbf{X}}}

\def\bbC{{\mathbb{C}}}

\def\bbR{{\mathbb{R}}}

\def\bbZ{{\mathbb{Z}}}

\def\cE{\mathcal{E}}
\def\cF{\mathcal{F}}

\def\cH{\mathcal{H}}

\def\cL{\mathcal{L}}

\def\cN{\mathcal{N}}
\def\cO{\mathcal{O}}

\def\cX{\mathcal{X}}

\def\sf0{{\mathsf{0}}}

\usepackage{epstopdf}
\usepackage{enumerate}
\usepackage{algorithm}
\usepackage{amsmath}
\usepackage{float}
\usepackage{color}
\usepackage{makeidx}
\usepackage{bm}
\usepackage{cleveref}
\usepackage{url}
\usepackage{steinmetz}
\usepackage{varwidth}
\usepackage{soul}
\usepackage{booktabs}

\DeclareMathOperator*{\argmax}{arg\,max}
\DeclareMathOperator*{\argmin}{arg\,min}

\usepackage{bbm}

\begin{document}
	\title{Millimeter Wave V2V Beam Tracking using Radar: Algorithms and Real-World Demonstration}
	
	\author{Hao~Luo,~Umut~Demirhan,~and~Ahmed~Alkhateeb\\School of Electrical, Computer, and Energy Engineering, Arizona State University\\Email: \{h.luo, udemirhan, alkhateeb\}@asu.edu
	}

	\maketitle
	
	\begin{abstract}
		Utilizing radar sensing for assisting communication has attracted increasing interest thanks to its potential in dynamic environments. A particularly interesting problem for this approach appears in the vehicle-to-vehicle (V2V) millimeter wave and terahertz communication scenarios, where the narrow beams change with the movement of both vehicles. To address this problem, in this work, we develop a radar-aided beam-tracking framework, where a single initial beam and a set of radar measurements over a period of time are utilized to predict the future beams after this time duration. Within this framework, we develop two approaches with the combination of various degrees of radar signal processing and machine learning. To evaluate the feasibility of the solutions in a realistic scenario, we test their performance on a real-world V2V dataset. Our results indicated the importance of high angular resolution radar for this task and affirmed the potential of using radar for the V2V beam management problems.
	\end{abstract}
	
	\section{Introduction}
	By employing massive antenna arrays and narrow beams, millimeter wave (mmWave) and terahertz (THz) communications can overcome challenging propagation conditions and leverage the extensive bandwidth offered in these frequency ranges to attain high data transfer speeds. This way, these bands can support the envisioned advanced vehicular technologies that demand high data rates \cite{va2016millimeter}. For example, critical safety applications can be further enhanced with the sharing of massive data generated from the vehicles equipped with a large number of sensors \cite{choi2016millimeter}. In such vehicular, and especially vehicle-to-vehicle (V2V) communication scenarios, achieving very high data rates requires accurately aligning the narrow beams adopted in these bands. Finding the optimal narrow beam, however, results in a large training overhead, which poses a major challenge in supporting highly-mobile vehicular scenarios. As the current and future channels are functions of the geometry of the environment and the position/direction of the transmitter/receiver, vehicular sensors of various modalities that are already available for other applications could be utilized to capture these features from the wireless environment. Especially with the integration of sensing and communication functions \cite{demirhan2022integrated}, the automotive radar sensors have become particularly interesting in aiding beam management in V2V scenarios. Hence, in this paper, we aim to investigate the use of radar sensing for beam tracking in V2V and evaluate the feasibility of the approach in the real world.
	
	Several real-world studies have been carried out to solve the beamforming with the aid of various sensing modalities in vehicle-to-infrastructure (V2I) scenarios, e.g., camera~\cite{jiang2022computer}, lidar~\cite{jiang2022lidar}, and radar~\cite{demirhan2022radarbeam}. 
	In~\cite{jiang2022computer}, the authors leverage the RGB images captured by the camera at the base station to eliminate the beam training overhead. Similarly, in~\cite{jiang2022lidar}, the lidar point cloud data of the communication environment is used to guide the beam prediction and tracking at the base station. More relevantly, in~\cite{demirhan2022radarbeam}, the beam prediction at the base station is conducted based on the radar sensing of the vehicular user. This work, however, was limited to a single radar target, which is highly limiting for V2V communication. For the V2V scenarios, \cite{aydogdu2020distributed} studied the radar-aided beamforming with the GPS data for the identification of the user. This work, however, focused on the initial access problem and utilized GPS data, which may be difficult to acquire, and relied on synthetic data.
	
	In this work, we aim to realize radar-aided beam tracking in V2V scenarios. For this purpose, we first formalize the radar-aided beam tracking problem by considering practical communication and radar models. In this problem, given an initial optimal beam and radar measurements from a longer duration, the purpose is to predict the optimal beam corresponding to the latest radar measurement. For this, we develop two LSTM-based approaches: (i) The radar data is first processed with classical methods to identify the radar state (i.e., range, angle, and Doppler) of the communication target, and then this information is utilized to predict the beam through a machine learning (ML) model. (ii) An end-to-end ML method using the radar maps along with the initial optimal beam. We evaluate the performance of the proposed solutions on real-world data collected with the V2V testbed of the DeepSense 6G dataset \cite{DeepSense,DeepSense_V2V}. Despite the limitations due to the low angular resolution of the radar, our results demonstrated the potential of using radar sensory data to aid the beam tracking for high data rate V2V communications.
	
	\section{System Model} 
\begin{figure}[!t]
		\centering
		\includegraphics[width=1\columnwidth]{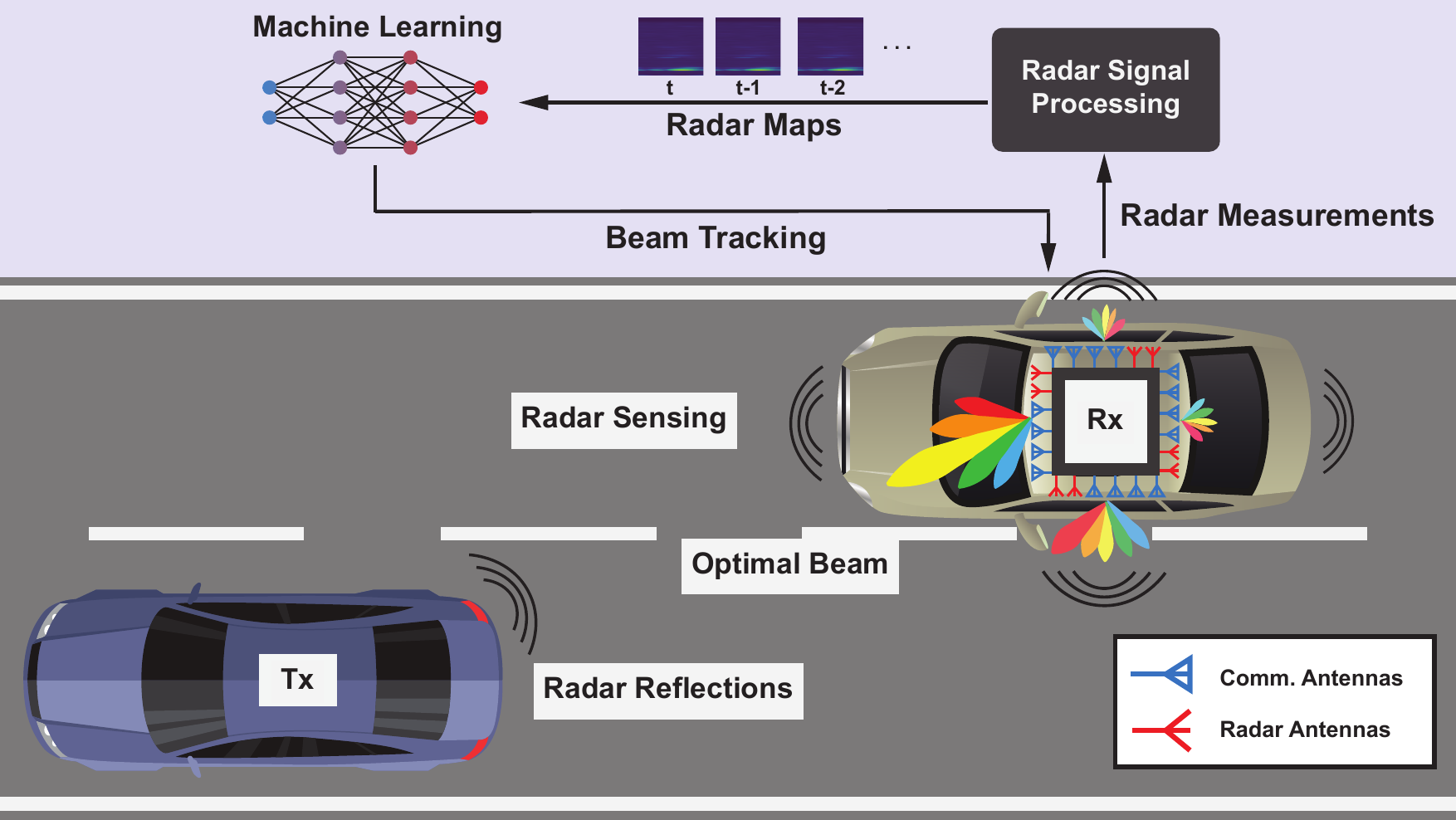}
		\caption{The figure presents the considered V2V system model. The receiver vehicle leverages the radar measurements to predict the optimal beam.}
		\label{fig:systemmodel}
	\end{figure}

	In this paper, a V2V communication scenario is considered, where the system model consists of a vehicle acting as the transmitter and another vehicle acting as the receiver, as illustrated in \figref{fig:systemmodel}.
	The transmitter employs a single omnidirectional antenna. Meanwhile, the receiver is equipped with (i) a set of linear analog mmWave antenna arrays, each directed towards different directions to cover the whole space, and (ii) a set of off-the-shelf FMCW radars operating at a different frequency band than the communication, each paired with one of the communication antenna arrays. To cover the whole circular directions, four linear mmWave antenna arrays are placed on the receiver vehicle in four separate directions, i.e., front, right, back, and left. In this model, the antenna arrays of the receiver are used to communicate with the transmitter. In this process, the mmWave beam selection is aided by the FMCW radars. 
	
	\subsection{Communication Model}
	In the communication model, the receiver employs four mmWave transceivers, each with $M_{a}^{c}$ antennas. Let $d \in \left\{ \textit{front}, \textit{right}, \textit{back}, \textit{left} \right\}$ be the communication and sensing direction of the receiver vehicle.
	With the geometric channel model~\cite{alkhateeb2014channel}, the channel between the transmitter and receiver in the direction $d$, $\bh_{d} \in \bbC^{M_{a}^{c}}$, can be given as 
	\begin{equation}
		\bh_{d} = \sum_{l=1}^{L_d} \alpha_{d, l} \ba(\theta_{d, l}^{az}, \theta_{d, l}^{el} ),
	\end{equation}
	where $L_d$ is the number of channel paths, and $\alpha_{d, l}$ is the complex gain.
	$\theta_{d, l}^{az}$ and $\theta_{d, l}^{el}$ are the azimuth and elevation angles of the paths, and $\ba(., .)$ is the array response vector of the antenna arrays.
	The transmitter transmits the data symbol $s$ to the receiver with the power gain $\cE_{c}$, and the receive array of direction $d$ receives the signal via the combining vector $\bff_{d} \in \bbC^{M_{a}^{c}}$.
	This received signal, $y_d$, can be written as
	\begin{equation}
		y_d = \sqrt{\cE_{c}} \bff_{d}^{H} \bh_{d} s + n,
	\end{equation}
	where $n \sim \cN(0, \sigma^{2})$ is the additive white Gaussian noise.
	The combining vector of the receive antenna array in the direction $d$, $\bff_{d}$, is assumed to be selected from a pre-defined codebook of $B$ beams, $\bm{\cF}_{d} = \left\{ \bff_{d, 1}, \ldots ,\bff_{d, B} \right\}$.
	Then, the indices of the optimal beam among the beams of these four antenna arrays and the corresponding antenna index can be represented as the result of the beamforming gain maximization given as
	\begin{gather}
		\begin{aligned}
			\max_{d, b} \quad & |\bff_{d, b}^{H} \bh_{d}|^{2} 
			\\
			\textrm{s.t.} \quad & d \in \{ \textit{front}, \textit{right}, \textit{back}, \textit{left} \}, \\
			& b \in \{1, \ldots, B\},
		\end{aligned}
	\end{gather}
	where the solution can be obtained by an exhaustive search. Note that the total number of combining vectors is $4 B$, with $B$ beams in each of the four directions.
	
	\subsection{Radar Model}
	To aid the communication, the FMCW radars on the receiver vehicle independently transmit the sensing signals and collect the reflected/scattered signals from the objects in the environment. Specifically, each FMCW radar transmits a sequence of $M_{\textrm{chirp}}$ chirps with a repetition interval of $T_{\textrm{PRI}}$ seconds, referred to as a radar frame. Let $s_{\textrm{chirp}}^{\textrm{tx}}(t) \in \bbR$ denote a single linear chirp with a duration of $T_{\textrm{active}}$ seconds and an instantaneous frequency $f_{0}+St$, where $S = \mathrm{BW}/T_{\textrm{active}} $ is the slope of the linear chirp with the bandwidth $\mathrm{BW}$. Then, the chirp signal $s_{\textrm{chirp}}^{\textrm{tx}}(t)$ can be written as \cite{li2021signal}
	\begin{equation}
		s_{\textrm{chirp}}^{\textrm{tx}}(t) = 
		\begin{cases}
			\sin(2\pi[f_{0}t + \frac{S}{2}t^{2}]) & \textrm{if} \ 0 \leq t \leq T_{\textrm{active}} \\
			0 & \textrm{otherwise},
		\end{cases}
	\end{equation}
	With this definition, the transmit signal of a radar frame, $s_{\textrm{frame}}^{\textrm{tx}}(t)$, can be written as
	\begin{equation}
		s_{\textrm{frame}}^{\textrm{tx}}(t) = \sqrt{\cE_{t}} \sum_{c=0}^{M_{\textrm{chirp}} - 1} s_{\textrm{chirp}}^{\textrm{tx}}(t - c \cdot T_{\textrm{PRI}}), \ 0 \leq t \leq T_{\textrm{frame}},
	\end{equation}
	where $\sqrt{\cE_{t}}$ is the transmission power gain, and $T_{\textrm{frame}}$ is the frame duration.
	
	At the radar receiver, the sensing signal reflected/scattered from the objects is first passed through a quadrature mixer.
	In the mixer, the received signal is mixed with two versions of the transmit signal $s_{\textrm{frame}}^{\textrm{tx}}(t)$, one with a $-90^\circ$ phase shift difference.
	Then, a low-pass filter is applied to the mixed signals to generate the intermediate frequency (IF) signal. 
	The IF signal is a constant frequency tone, which reflects the difference in the instantaneous frequency of the transmit and receive chirp signals.
	If a single object is in front of the radar, the IF signal of a chirp can be expressed as
	\begin{equation}
		s_{\textrm{chirp}}^{\textrm{rx}}(t) = \sqrt{\cE_{t} \cE_{r}} \exp\left(j 2 \pi \left[ S \tau t + f_{0}\tau - \frac{S}{2}\tau^{2} \right] \right),
	\end{equation}
	where $\cE_{r}$ is the reflection/scattering gain consisting of the radar cross section (RCS) and the path-loss.
	$\tau= R/ \varsigma$ is the round-trip time of the sensing signal, where $R$ is the total propagation distance, and $\varsigma$ is the speed of light.
	
	Finally, the received IF signals of the chirps are sampled by an analog-to-digital converter (ADC) with the sampling rate $F_{s}$, where each chirp is sampled with $M_{\textrm{sample}}$ samples.
	Assuming the FMCW radar has $M_{\textrm{ant}}$ receive antennas, and each antenna has its own RF receive chain, the measurements of one radar frame at the radar in the direction $d$ can be denoted by $\bX_d \in \bbC^{M_{\textrm{ant}} \times M_{\textrm{sample}} \times M_{\textrm{chirp}}}$.
	
	\section{Problem Definition}
	In this work, we aim to predict the optimal beam for the receiver vehicle based on the recent radar measurements from the surrounding environment. Specifically, after communication is established, the communication beam in the mobile environment may need to be frequently updated. To that end, tracking the transmitter vehicle within the radar measurements and then updating the optimal beam without additional beam measurements can reduce the resources needed for beam management and, particularly, beam tracking. For the beam tracking, however, using only the radar measurements is not sufficient since the transmitter vehicle needs to be identified in the radar. For this purpose, in addition to the radar measurements, we leverage the initial optimal beam index in our problem, which could be estimated/becomes available during the initial establishment of the communication.
	
	The radar-aided beam tracking in the V2V scenario is a challenging task due to (i) multiple objects in the highly dynamic environment, (ii) noisy radar data from the mobile receiver/radar, and (iii) multiple potential directions of linear arrays. To simplify (iii), we only focus on the tracking within a single receive array/radar pair and assume that the receive array/radar pair does not change within the sequence of samples that have been tracked. Thus, we can drop the sub-index $d$, which indicates the direction of the radar and phased array. It is worth noting that, however, this operation induces additional difficulty for the proposed algorithm since it needs to accommodate the data from different receiver/radar pairs.
	
	To formalize the problem, we first denote the latest $T_o \in \bbZ^{+}$ (referred to as the observation interval) radar measurements at time $t$ by $\bm{\cX}_{t} = \{ \bm{\mathrm{X}}_{t-T_{o}+1}, \ldots, \bm{\mathrm{X}}_{t} \}$. We then define the function $f_{\bm{\Theta}}$ of parameters $\boldsymbol{\Theta}$, that maps the $T_o$ latest radar measurements, $\bm{\cX}_{t}$, and the optimal beam index of the first radar measurement, $b_{t-T_{o}+1}^{*}$, to the current optimal beam index, $b_{t}^{*}$. This function can be expressed as $f_{\bm{\Theta}}(\bm{\cX}_{t}, b_{t-T_{o}+1}^{*}) = b_{t}^{*}$. Now, our objective is to optimize the design of the mapping function and the set of parameters $\bm{\Theta}$ in order to maximize the predicted beam's accuracy, which can be written as
	\begin{equation}
		\hat{f}_{\hat{\bm{\Theta}}} = \argmax_{f, \bm{\Theta}} \frac{1}{T} \sum_{t=1}^{T} \bm{1}_{\left\{ b_{t}^{*} = f_{\bm{\Theta}}(\bm{\cX}_{t}, b_{t-T_{o}+1}^{*}) \right\}},
	\end{equation}
	where $T$ is the total number of data samples, and $\bm{1}_{E}$ is the indicator function of the event $E$, i.e., $\bm{1}_{E}=1$ if $E$ occurs; otherwise, $\bm{1}_{E}=0$.
	
	\section{Proposed Solutions}
	For the defined radar-aided V2V beam tracking problem, we develop two LSTM-based beam tracking approaches: (i) The LSTM is fed with the transmitter vehicle's radar state estimation  (range, angle, and Doppler) to track the beams, and (ii) an end-to-end ML approach where the radar maps are directly fed to the LSTM and the output is combined with the previous optimal beam index to obtain the tracked beam. Note that both solutions adopt a certain level of classical radar processing, and hence, in the following, we first introduce the radar preprocessing methods and then present our solutions.
	
	\textbf{Radar preprocessing:} To extract useful information from the radar measurements, e.g., range, angle, and Doppler, we adopt conventional signal processing techniques. In particular, we apply FFTs to generate the range-Doppler maps and the radar cube (i.e., range-angle-Doppler maps). We use both maps for the state estimation of the objects in the first method and the range-Doppler maps as input in the second end-to-end method. Mathematically, we describe the range-angle maps and the radar cube as
	\begin{equation}
		\bm{\mathrm{H}}^{\textrm{RD}} = \sum_{a=1}^{M_{\textrm{ant}}}|\cF_{\mathrm{2D}}(\bm{\mathrm{X}}_{a,:,:})|, \quad
		\bm{\mathrm{H}}^{\textrm{RC}} = |\cF_{\mathrm{3D}}(\bm{\mathrm{X}})|,
	\end{equation}
	where $\cF_{\mathrm{2D}}(.)$, and $\cF_{\mathrm{3D}}(.)$ denote the 2D and 3D FFT operations. For more detailed information on these maps, please refer to \cite{demirhan2022radarbeam, li2021signal}.
	
	\begin{figure}[!t]
		\centering
		\includegraphics[width=0.9\columnwidth]{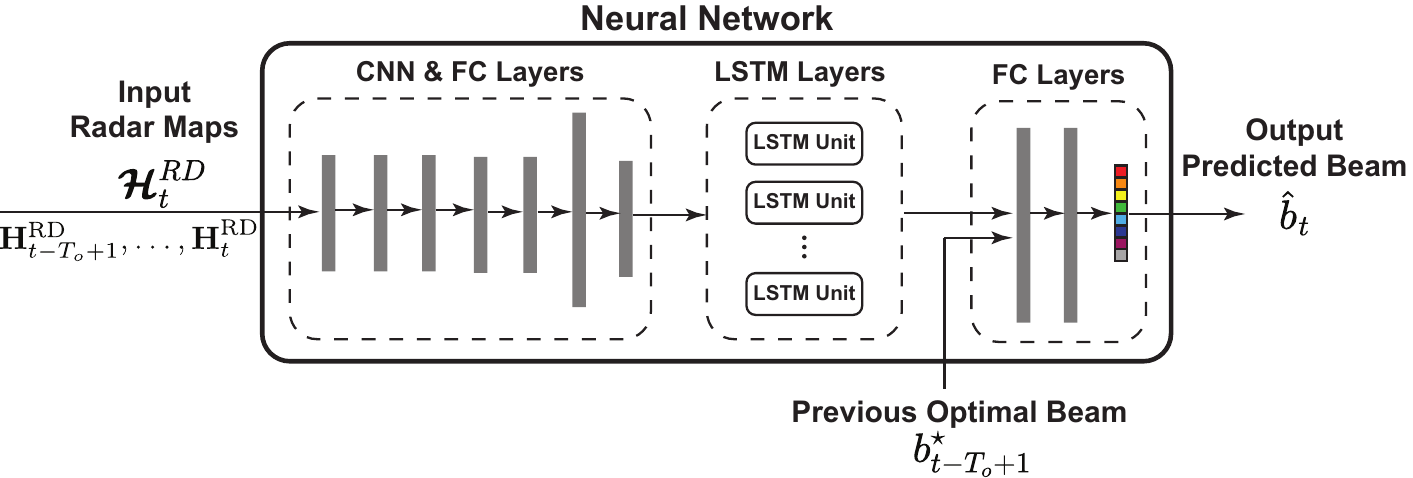}
		\caption{The illustration of the end-to-end model. In this model, the range-Doppler maps are fed to the shown architecture. The initial optimal beam is used to provide the historical information of the transmitter.}
		\label{fig:end_to_end}
	\end{figure}
	
	\subsection{Beam Tracking with Transmitter Identification}
	In this approach, given a sequence of radar measurements $\bm{\cX}_{t}$ and a previous optimal beam index $b_{t-T_{o}+1}^{*}$, we first detect and then track the transmitter vehicle over the radar samples with a signal processing based approach. After that, we use an LSTM model to predict the current beam based on the information of the transmitter vehicle among the time samples.
	
	\textbf{Object detection:} For the detection of the objects in each radar sample, we apply a classical radar detection chain. Specifically, we (i) obtain the range-Doppler map and radar cube for each radar measurement, (ii) apply a CFAR method to each range-Doppler map to detect the points with high reflection/scattering power, (iii) cluster the detected points using DBSCAN to determine the objects, (iv) estimate the angle from the object's range and Doppler slice in the radar cube, where the angle bin with the peak power is taken as the estimate. Then, for a sequence of radar measurements, we have a sequence of detected objects $\bm{\cO}_{t}=\{ \bm{\mathrm{O}}_{t-T_{o}+1}, \ldots, \bm{\mathrm{O}}_{t} \}$, where $\bm{\mathrm{O}}_t=\{ \bm{\mathrm{o}}_{t, 1}, \ldots, \bm{\mathrm{o}}_{t, K_{t}}  \}$ denotes the $K_{t}$ objects detected in the radar measurement at time $t$, and $\bm{\mathrm{o}}_{t, k}=\{ r_{t,k}, v_{t,k}, a_{t,k} \}$ denotes the state of $k$-th object with range $r_{t,k}$, Doppler velocity $v_{t,k}$, and angle $a_{t,k}$.
	
	\begin{figure*}[!t]
		\centering
		\includegraphics[width=1.9\columnwidth]{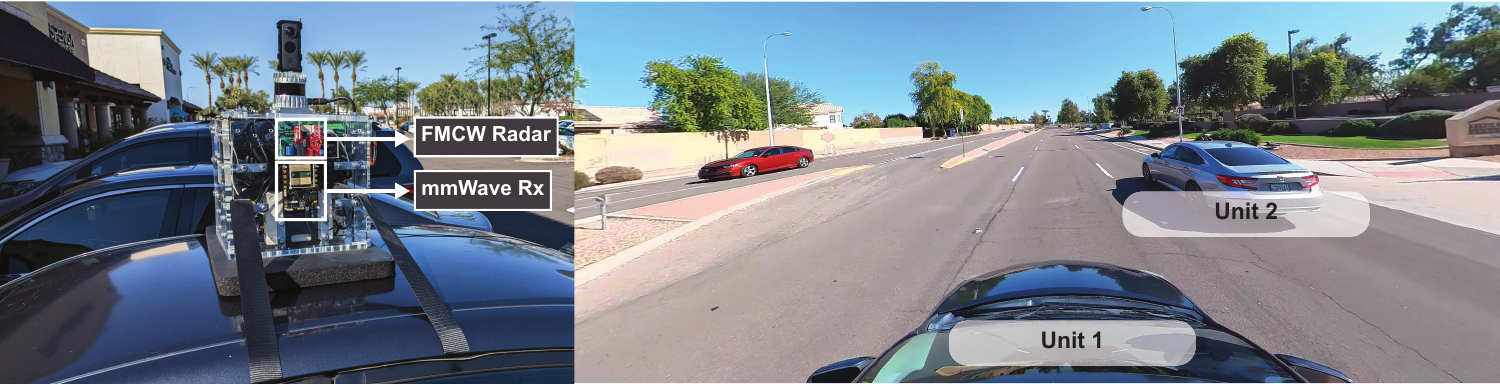}
		\caption{The image on the left depicts the setup of the DeepSense 6G dataset's Testbed 6, where mmWave antenna arrays and radars are placed on top of a vehicle (Unit 1). Four sets of mmWave receivers and radars face four sides of the box. The image on the right shows a sample from the collection.}
		\label{fig:testbed}
	\end{figure*}
	
	\textbf{Transmitter identification and tracking:} Next, we need to determine the transmitter in the radar samples. For this purpose, we develop a communication angle-based method. Let $\psi_{t,k}^{r}=a_{t,k}$ and $\psi_{t}^{c}$ denote the radar angle of the $k$-th object and the angle corresponding to the optimal beam at time $t$, respectively. To identify the transmitter, we select the object that has the radar angle closest to the communication angle, i.e., $\hat{k}_{t} = \argmin_{k \in \{ 1, \cdots, K_{t}\}} |\psi_{t,k}^{r} - \psi_{t}^{c}|$,
	which can be used to track the transmitter in the following radar samples. Specifically for the tracking, we select the object that has the closest distance to the previously determined transmitter state with the following metric
	\begin{equation}
		\hat{k}_{t+1} = \argmin_{k} w_r \left| r_{t+1,k} - r_{t, \hat{k}_{t}}\right| + w_v \left| v_{t+1,k} - v_{t, \hat{k}_{t}}\right|,
	\end{equation}
	where $w_r$ and $w_v$ are the weights. After applying the tracking process iteratively, we obtain the information of the transmitter through a sequence of radar measurements, denoted by $\bm{\mathrm{O}}_{t}^{\textrm{tx}}=\{ \bm{\mathrm{o}}_{t-T_{o}+1}^{\textrm{tx}}, \ldots, \bm{\mathrm{o}}_{t}^{\textrm{tx}} \}$, which is ready to be utilized in the ML.
	
	\noindent\textbf{Beam tracking:} As the proposed detection chain may present errors in the highly dynamic environment, we adopt the deep learning to complement it with a robust model for the tracking of the transmitter and beams. In this model, the LSTM takes the extracted radar information of the transmitter over time as the input and returns the output to a set of two fully connected layers to obtain the beam. This output is of $\mathbb{R}^B$, where each element corresponds to a beam index. To optimize the parameters of the model, $\bm{\Theta}$, we use the loss between the output of the networks and the one-hot encoding of the optimal beam index given as
	\begin{equation} \label{eq:loss}
		\bm{\Theta}^{*} = \argmin_{\Theta} \frac{1}{T} \sum_{t=1}^{T} \cL (g_{\boldsymbol{\Theta}}(\bm{\mathrm{O}}_{t}^{\textrm{tx}}), b_{t}^{*}),
	\end{equation}
	where $g_{\boldsymbol{\Theta}}(.)$ is the neural network function, and $\cL$ is the cross-entropy loss. Next, we develop an end-to-end solution that does not rely on object detection and transmitter identification.
	
	\subsection{End-to-End Machine Learning Approach}
	For the end-to-end solution, we aim to predict the beam directly using the range-Doppler maps and the previous optimal beam index. For this purpose, we use a deep learning model with convolutional, LSTM, and fully connected layers, as illustrated in \figref{fig:end_to_end}. Before the LSTM, the range-Doppler maps of different time samples are processed by the same five convolutional layers with the rectified linear unit (ReLU) activations and average pooling.
	The output of the fifth convolutional layer for each time sample is then connected to the corresponding LSTM unit. The output of the LSTM is combined with the given previous beam information and fed to three fully connected layers to return the beam. If we denote $\bm{\cH}_{t}^{\textrm{RD}}=\{ \bm{\mathrm{H}}_{t-T_{o}+1}^{\textrm{RD}},\ldots,\bm{\mathrm{H}}_{t}^{\textrm{RD}} \}$ as the sequences of range-Doppler maps, we can write the objective of the model as
	\begin{equation}
		\bm{\Theta}^{*} = \argmin_{\Theta} \frac{1}{T} \sum_{t=1}^{T} \cL (g_{\Theta}(\bm{\cH}_{t}^{\textrm{RV}}, b_{t-T_{o}+1}^{*}), b_{t}^{*}).
	\end{equation}
	Even though the method developed in this subsection includes less classical radar signal processing, it may suffer from the difficulty of learning very complicated processing (object detection and transmitter identification) without a huge complexity and number of samples.

	\section{Dataset}
	For a realistic evaluation of the proposed beam tracking solutions, we utilize a real-world dataset that is collected by a V2V testbed as a part of the DeepSense 6G \cite{DeepSense,DeepSense_V2V} framework, with co-existing radar and communication equipment.

	\textbf{Testbed:} 
	We adopt the Testbed 6 of the DeepSense 6G dataset~\cite{DeepSense}, and the setup of the testbed is presented in~\figref{fig:testbed}.
	There are two units in this testbed: (i) Unit 1, a mobile receiver that has four pairs of an FMCW radar and a 60GHz mmWave receiver facing four directions, i.e., front, right, back, left, and (ii) Unit 2, a mobile transmitter that uses a 60 GHz quasi-omni antenna.
	Each mmWave receiver adopts a uniform linear array (ULA) with $M_{a}^{c}=16$ elements and an over-sampled beamforming codebook of $B=64$ vectors.
	Also, the mmWave receivers apply their codebook beams as a combiner to capture the received power, and the combiner having the strongest power is taken as the optimal beam.
	For the radar, only one of the transmit antennas and $M_{\textrm{ant}}=4$ receive antennas are activated. We adopt a set of radar parameters, given by chirp starting frequency $f_{0}=77$ GHz, ADC sampling rate $F_{s}=5$ us, chirp slope $S=15$ MHz/us, $M_{chirp}=128$ chirps/frame, and $M_{sample}=256$ samples/chirp. This setting provides the bandwidth $\textrm{BW}=768$ MHz, a maximum range of 50 meters, and a maximum velocity of 54 km/hr.
	
	\textbf{AI-Ready Dataset:} 
	For the evaluation, we present Scenario 38 of the DeepSense 6G dataset. In this scenario, the transmitter was fixed on a tripod extending to the sunroof of the vehicle (Unit 2) while the two vehicles traveled within the traffic in close proximity of each other. The samples are collected continuously with a sampling rate of $10$ samples/s, and they include a variety of traffic scenarios, including following, passing, and changing lanes. To focus on the scenarios interesting for tracking, we filter the data by keeping the sequences of samples with changing beam indices (e.g., the transmitter getting closer to the receiver in a different lane or the vehicles consecutively taking a turn). Also, for the samples of beam tracking, we generate overlapping data sequences with a length of $10$. For the shorter observation interval, we only utilized the last $T_o\leq10$ samples of each pre-generated longer sequence. The resulting final dataset of 3649 sequences is split into the training and test set with a 70/30\% ratio. To prevent overfitting in the test set, the training and test samples are selected from different (non-overlapping) time intervals.

	\section{Results}
	In this section, we present the performance of the proposed solutions to evaluate the feasibility of the radar-aided V2V beam tracking problem and proposed approaches in the real world. 
	For the training of (i) the transmitter identification-based solution, we train the network for 80 epochs using the Adam algorithm with a learning rate 0.01, batch size 32, and a decay factor $\gamma=0.01$ applied every 20 epochs. For (ii) the end-to-end learning solution, we adopt a similar set of parameters with a learning rate 0.001 and the decay factor $\gamma=0.1$ applied every 40 epochs.
	
	\begin{figure}[!t]
		\centering
		\includegraphics[width=.75\columnwidth]{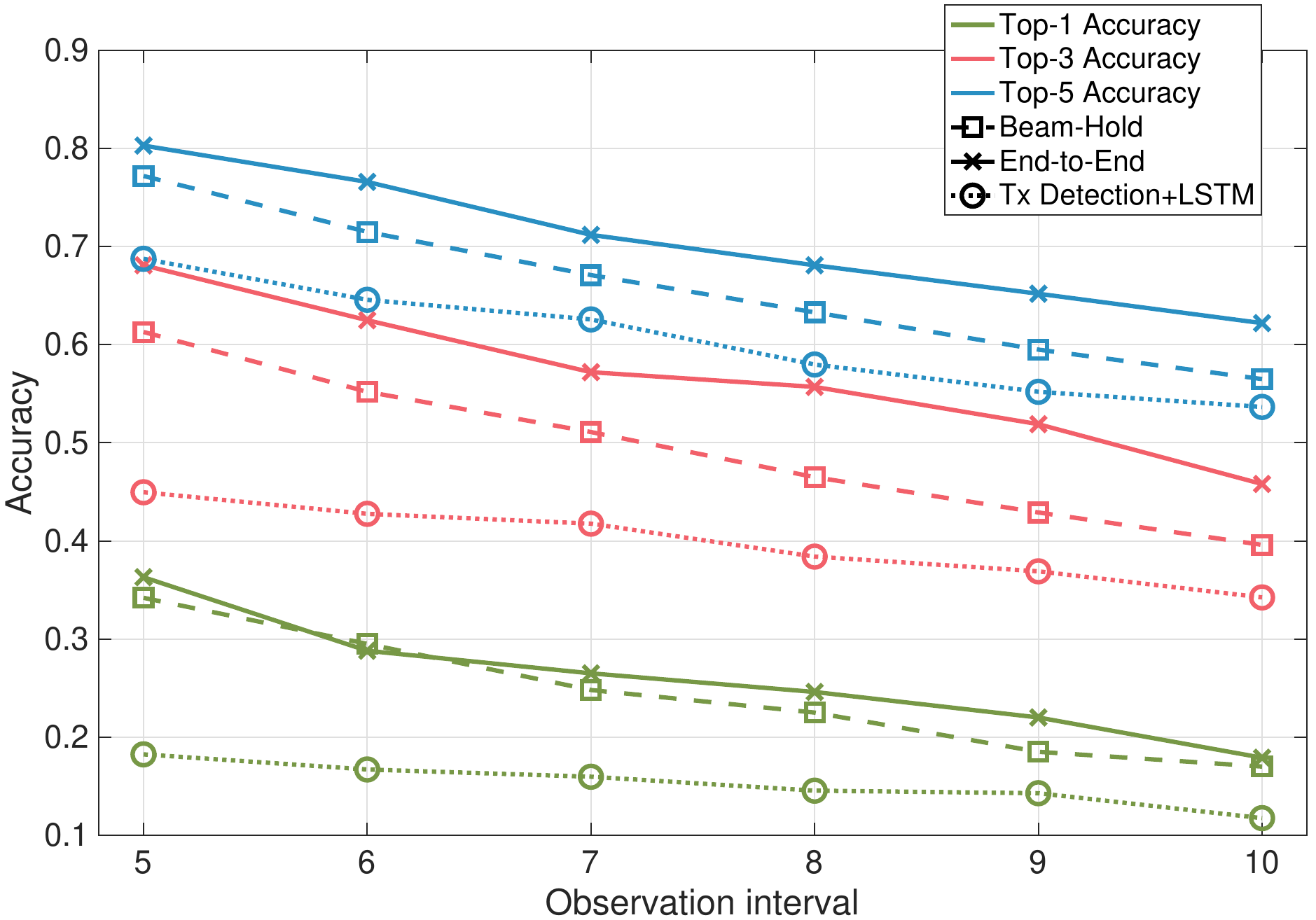}
		\caption{The top-k test accuracy of the proposed approaches. The proposed solutions show less performance degradation for longer intervals compared to beam hold, showing the advantage of radar-aided beam tracking.}
		\label{fig:acc_vs_duration}
	\end{figure}
	
	For the evaluation, we utilize the top-$k$ accuracy of the beam tracking solutions. As the baseline, we adopt a beam hold method, where the previous beam given as input is used as the predicted beam, i.e., $\hat{b}_t={b}^\star_{t-T_o+1}$. For the top-$3$ and -$5$ predictions of this baseline solution, $\pm 1$ and $\pm 2$ beams are utilized. This baseline allows evaluation of the radar data's effect. In \figref{fig:acc_vs_duration}, we present the performance of the solutions for various numbers of radar samples ($T_o$). Interestingly, the transmitter identification-based approach shows worse performance than the baseline. The difference, however, decreases with the observation interval and higher $k$ values. \textit{This shows the limitation due to the low angular resolution of the radar compared to the communication antennas/beams, where the exact prediction of the beam directly from the radar is difficult.} By contrast, the end-to-end solution overall outperforms the baseline. \textit{While the gain is limited for top-$1$ accuracy, it provides $6$-$9\%$ and $3$-$6\%$ gain for the top-$3$ and -$5$ accuracies and shows the potential of radar-aided beam tracking.}
	
\begin{figure}[!t]
		\centering
		\subfigure[Transmitter Identification]{\label{fig:cm_tx_id}\includegraphics[width=0.49\columnwidth]{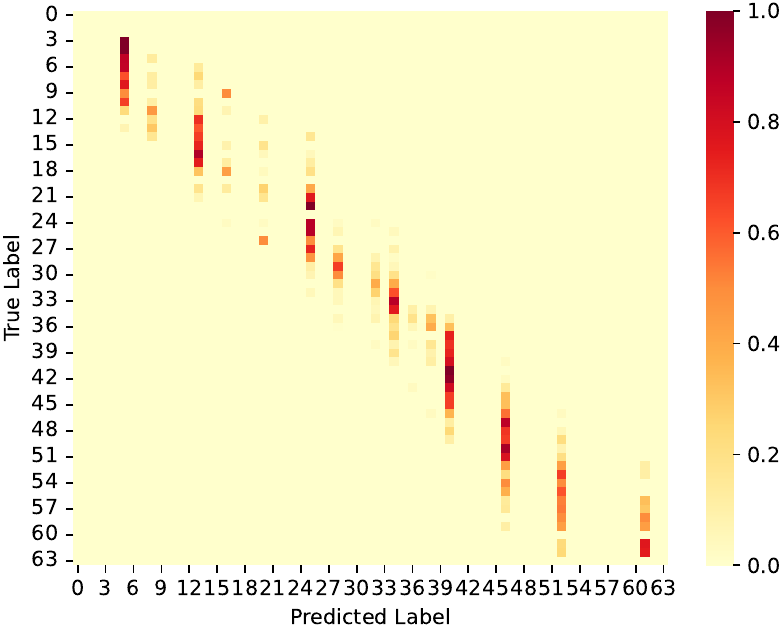}}
		\hfill
		\subfigure[End-to-end Learning]{\label{fig:cm_e2e}\includegraphics[width=0.49\columnwidth]{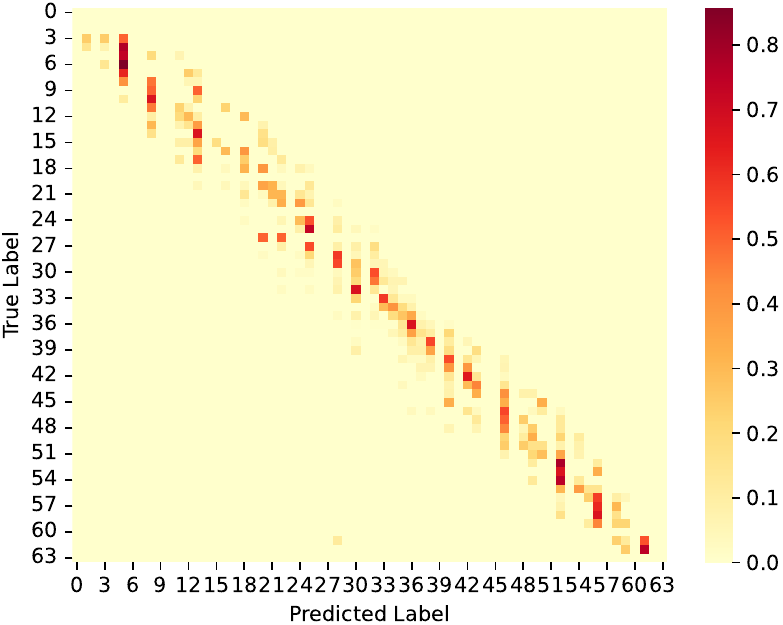}}
		\caption{The confusion matrix of the predictions for $T_o=5$ radar samples. The low angular resolution results a bias toward specific bins as shown in (a).}
		\label{fig:confusion_matrix}
	\end{figure}

	In \figref{fig:confusion_matrix}, we illustrate the confusion matrix of the solutions. \figref{fig:cm_tx_id} again shows the limitation of the low angular resolution of the radar, which causes a bias towards specific bins with the radar target tracking approach. On the contrary, \figref{fig:cm_e2e} shows a linear pattern with higher accuracy, where the given beam is refined with the radar information. \textit{To conclude, although high angular resolution in radar may be required to achieve the full potential, the aid of the radar can help to overcome beamforming problems in very challenging real-world V2V communication applications.}

	\section*{Acknowledgement}
	This work is supported in part by the National Science Foundation under Grant No. 2048021.

\end{document}